\documentclass[11pt,english,aip,floatfix,reprint,sort&compress]{revtex4-1}

\usepackage[T1]{fontenc}
\usepackage[latin9]{inputenc}
\usepackage[letterpaper]{geometry}
\usepackage{color}
\usepackage{babel}
\usepackage{array}
\usepackage{float}
\usepackage{multirow}
\usepackage{amsmath}
\usepackage{amssymb}
\usepackage{graphicx}
\usepackage[unicode=true,
 bookmarks=true,bookmarksnumbered=false,bookmarksopen=false,
 breaklinks=false,pdfborder={0 0 1},backref=false,colorlinks=false,pdfpagemode=FullScreen]
 {hyperref}
\hypersetup{
 pdfauthor={Clementi}}

\makeatletter

\providecommand{\tabularnewline}{\\}
\floatstyle{ruled}
\newfloat{algorithm}{H}{loa}
\providecommand{\algorithmname}{Algorithm}
\floatname{algorithm}{\protect\algorithmname}

\@ifundefined{date}{}{\date{}}
\usepackage{xr}
\makeatother

\begin{document}

\title{Ensemble Learning of Coarse-Grained Molecular Dynamics Force Fields
with a Kernel Approach}

\author{Jiang Wang{*}}

\address{Rice University, Center for Theoretical Biological Physics, Houston,
Texas 77005, United States}

\address{Rice University, Department of Chemistry, Houston, Texas 77005, United
States}

\author{Stefan Chmiela{*}}

\address{Technische Universität Berlin, Machine Learning Group, 10587 Berlin,
Germany}

\author{Klaus-Robert Müller}

\address{Technische Universität Berlin, Machine Learning Group, 10587 Berlin,
Germany}

\address{Korea University, Department of Brain and Cognitive Engineering,
Seoul, 02841, Republic of Korea}

\address{Max Planck Institute for Informatics, Saarbrücken, 66123, Germany}

\author{Frank Noé}
\email{frank.noe@fu-berlin.de}

\selectlanguage{english}%

\address{Freie Universität Berlin, Department of Mathematics and Computer
Science, Arnimallee 6, 14195 Berlin, Germany}

\address{Rice University, Center for Theoretical Biological Physics, Houston,
Texas 77005, United States}

\address{Rice University, Department of Chemistry, Houston, Texas 77005, United
States}

\address{Freie Universität Berlin, Department of Physics, Arnimallee 14, 14195
Berlin, Germany}

\author{Cecilia Clementi}
\email{cecilia@rice.edu}

\selectlanguage{english}%

\address{Rice University, Center for Theoretical Biological Physics, Houston,
Texas 77005, United States}

\address{Rice University, Department of Chemistry, Houston, Texas 77005, United
States}

\address{Rice University, Department of Physics, Houston, Texas 77005, United
States}

\address{Freie Universität Berlin, Department of Mathematics and Computer
Science, Arnimallee 6, 14195 Berlin, Germany}

\address{Freie Universität Berlin, Department of Physics, Arnimallee 14, 14195
Berlin, Germany}
\begin{abstract}
Gradient-domain machine learning (GDML) is an accurate and efficient
approach to learn a molecular potential and associated force field
based on the kernel ridge regression algorithm. Here, we demonstrate
its application to learn an effective coarse-grained (CG) model from
all-atom simulation data in a sample efficient manner. The coarse-grained
force field is learned by following the thermodynamic consistency
principle, here by minimizing the error between the predicted coarse-grained
force and the all-atom mean force in the coarse-grained coordinates.
Solving this problem by GDML directly is impossible because coarse-graining
requires averaging over many training data points, resulting in impractical
memory requirements for storing the kernel matrices. In this work,
we propose a data-efficient and memory-saving alternative. Using ensemble
learning and stratified sampling, we propose a 2-layer training scheme
that enables GDML to learn an effective coarse-grained model. We illustrate
our method on a simple biomolecular system, alanine dipeptide, by
reconstructing the free energy landscape of a coarse-grained variant
of this molecule. Our novel GDML training scheme yields a smaller
free energy error than neural networks when the training set is small,
and a comparably high accuracy when the training set is sufficiently
large.
\end{abstract}
\maketitle

\section{Introduction}

Molecular dynamics (MD) simulations have become an important tool
to characterize the microscopic behavior of chemical systems. Recent
advances in hardware and software allow significant extensions of
the simulation timescales to study biologically relevant processes
\citep{LindorffLarsenEtAl_Science11_AntonFolding,BuchEtAl_JCIM10_GPUgrid,ShirtsPande_Science2000_FoldingAtHome}.
For example, we can now characterize the configurational changes,
folding and binding behavior of small to intermediate-sized proteins
through MD on the timescale of milliseconds to seconds \citep{DrorEtAl_PNAS11_DrugBindingGPCR,ShuklaPande_NatCommun14_SrcKinase,PlattnerNoe_NatComm15_TrypsinPlasticity,PlattnerEtAl_NatChem17_BarBar,PaulEtAl_PNAS17_Mdm2PMI,NoeAnnRev2020}.
However, the computational complexity of evaluating the potential
energy prohibits this approach to scale up to significantly larger
systems and/or longer timescales. Therefore, multiple ways have been
proposed to speed up atomistic simulations, such as advanced sampling
methods (e.g., umbrella sampling \citep{Frenke2001un,Torrie1977,Johannes2011},
parallel tempering \citep{Swendsen1986,Neal1996}) or adaptive sampling
\citep{Rajamani_Proteins24_1775,PretoClementi_PCCP14_AdaptiveSampling,BowmanEnsignPande_JCTC2010_AdaptiveSampling}.
An alternative approach is to reduce the dimensionality of the system
by coarse-graining (CG) \citep{ClementiCOSB,Davtyan2012,Izvekov2005,Marrink2004,MllerPlathe2002,Noid2013}.
The fact that macromolecules usually exhibit robust collective behavior
suggests that not every single degree of freedom is per se essential
in determining the important macromolecular processes over long timescales.
Furthermore, a CG representation of the system simplifies the model
and allows for a more straightforward physico-chemical interpretation
of large-scale conformational changes such as protein folding or protein-protein
binding\citep{ClementiCOSB}.

Once the mapping from the atomistic to the CG representation is defined,
a fundamental challenge is the definition of an effective potential
in reduced coordinates, such that the essential physical properties
of the system under consideration are retained. The choice of the
relevant properties crucially dictates the definition of the CG model.

Following a top-down approach, the CG procedure is driven by the objective
to reproduce macroscopic properties, such as structural information
or experimentally measured observables \citep{Nielsen2003,MatysiakClementi_JMB04_Perturbation,MatysiakClementi_JMB06_Perturbation,Marrink2004,Davtyan2012,Chen2018}.
In a bottom-up approach, on the other hand, an effective potential
is designed to reproduce a selection of properties of an atomistic
model, for instance the probability distribution in a suitable space
and the corresponding metastable states \citep{Lyubartsev1995,MllerPlathe2002,Praprotnik2008,Izvekov2005,Wang2009,Shell2008}.

In the past several years, machine learning (ML) techniques have been
increasingly applied in molecular simulation \citep{noe2019science,NOE202077,NoeAnnRev2020,Schutt2019}.
Bottom-up CG methods have also started to leverage the advances in
ML, to define classical atomistic potentials or force fields from
quantum chemical calculations \citep{BehlerParrinello_PRL07_NeuralNetwork,Bartok2010,Rupp2012,Bartok2013,Smith2017,Bartok2017,Schuett2017,Smith2018,Schuett2018,Grisafi2018,Imbalzano2018,Nguyen2018,ZhangHan2018,ZhangHan2018_PRL,BereauEtAl_JCP18_Physical,ChmielaEtAl_SciAdv17_EnergyConserving,ChmielaEtAl_NatComm18_TowardExact},
to learn kinetic models \citep{MardtEtAl_VAMPnets,WuEtAl_NIPS18_DeepGenMSM,WehmeyerNoe_TAE,HernandezPande_VariationalDynamicsEncoder,RibeiroTiwary_JCP18_RAVE},
or to design effective CG potential from atomistic simulations \citep{John2017,ZhangHan2018_CG,wang2019}.
In this context, we have recently shown that a deep neural network
(NN) can be used in combination with the well established ``force
matching'' approach \citep{Noid2008} to define a coarse-grained
implicit water potential that is able to reproduce the correct folding/unfolding
process of a small protein from atomistic simulations in explicit
water \citep{wang2019}. In the force matching approach, the effective
energy function of the CG model is optimized variationally, by finding
the CG force field that minimizes the difference with the instantaneous
atomistic forces projected on the CG coordinates. As there are multiple
atomistic configurations consistent with a CG configuration, this
estimator is very noisy and this approach requires a large amount
of training data. It is thus restricted to parametric models like
NNs, as the computational complexity of non-parametric models is directly
linked to training set size. Here we propose a method to overcome
this limitation in the dataset via bootstrap aggregation in combination
with a non-parametric, kernel-based regressor.

In particular, we use the Gradient-Domain Machine Learning (GDML)
approach \citep{ChmielaEtAl_SciAdv17_EnergyConserving,CHMIELA201938}.
In the application to quantum data, GDML is able to use a small number
(usually less than a few thousands) of example points to build an
accurate force field for a specific molecule. Because of the degeneracy
of the mapping, the training data required to reconstruct a coarse-grained
force field is much larger, and the fact that memory requirements
scale quadratically with data set size prevents a direct application
of GDML to the definition of CG models.

To solve this problem, we pursue a hierarchical ensemble learning
approach in which the full training set is first divided into smaller
batches that are trained independently. A second GDML layer is then
applied to the mean prediction of this ensemble, providing the second
model with a consistent set of inputs and outputs. We show that GDML
with ensemble learning can be efficiently used for the coarse-graining
of molecular systems.

The structure of the paper is as follows. In the section ``Theory
And Methods'', we briefly review the principle of force matching
that we use for coarse graining, as well as mathematical underpinnings
of the kernel ridge regression used in the GDML method. Then we describe
the idea of ensemble learning, and explain how it solves the problem
associated with the large number of training points required by force
matching. In the ``Results'' section, we demonstrate that a GDML
approach trained with ensemble learning performs well on the coarse-graining
of a small molecular system, alanine dipeptide simulated in water
solvent, as it produces the same free energy surface as obtained in
the all-atom simulations. As it was already demonstrated in the case
of a NN approach \citep{wang2019}, the key to success of a GDML-based
coarse graining is that it is able to naturally capture nonlinearities
and multi-body effects arising from the renormalization of degrees
of freedom at the base of coarse-graining.

\section{Theory and methods}

\subsection{Coarse-Graining with Thermodynamic Consistency}

Although the definition of a coarse-graining mapping scheme is per
se an interesting problem \citep{BoninsegnaBanish2018,sinitskiy2012optimal,Noid2013,WangBombarelli2019},
here we start by assuming that a mapping is given. The all-atom system
we want to coarse-grain consists of $N$ atoms, and its configurations
are represented by a $3N$ dimensional vector $\mathbf{r}\in\mathbb{R}^{3N}$.
The lower dimensional CG representation of the system is given by
the mapping:
\begin{equation}
\mathbf{x}=\xi(\mathbf{r})\in\mathbb{R}^{3n}\label{eq:CGmapping}
\end{equation}

where $n<N$ is the number of CG beads. The CG mapping function $\xi$
is assumed to be linear, i.e. there exists a coarse-graining matrix
$\Xi\in\mathbb{R}^{3n\times3N}$ that maps the all atom space to the
CG space: $\mathbf{x}=\Xi\mathbf{r}$.

The definition of a CG model requires an effective potential $U(\mathbf{x};\boldsymbol{\theta})$
in the CG space, where $\theta$ are the optimization parameters.
The potential $U(\mathbf{x};\boldsymbol{\theta})$ can then be used
to generate an MD trajectory with a dynamical model. Parameterizations
are available in varying degrees of sophistication, ranging from classical
force fields with fixed functional forms, to ML approaches with strong
physical basis.

One popular bottom-up method for building a CG model is to require
thermodynamic consistency, that is to design a CG potential such that
its equilibrium distribution matches the one of the all-atom model.
In practice, this means that an optimum CG potential satisfies the
condition:

\begin{equation}
U(\mathbf{x};\boldsymbol{\theta})\equiv-k_{B}T\ln p^{CG}(\mathbf{x})+\mathrm{const},\label{eq:free-ene}
\end{equation}
where $k_{B}$ is the Boltzmann constant, $T$ is the temperature,
and the probability density distribution in the CG space is given
by the equilibrium distribution of the all-atom model mapped to the
CG coordinates:
\begin{equation}
p^{CG}(\mathbf{x})=\frac{\int\mu(\mathbf{r})\delta\left(\mathbf{x}-\xi(\mathbf{r})\right)d\mathbf{r}}{\int\mu(\mathbf{r})d\mathbf{r}}\label{eq:prob}
\end{equation}
where $\text{\ensuremath{\mu}(\ensuremath{\mathbf{r}})=\ensuremath{\exp\left(-V(\mathbf{r})/k_{B}T\right)}}$
is the Boltzmann weight associated with the atomistic energy $V(\mathbf{r})$.

Different methods have been proposed to construct a CG potential $U(x,\theta)$
that satisfy Eq.\ref{eq:prob}, notably the relative entropy method
\citep{Shell2008}, and the force-matching method\citep{Izvekov2005,Noid2008}.
In this work we will demonstrate how we could learn the molecular
CG potential using the idea of force-matching and the GDML kernel
method.

\subsection{Force Matching}

It can be shown that the potential $U(\mathbf{x};\boldsymbol{\theta})$
satisfies thermodynamic consistency if the associated CG forces $-\nabla U(\mathbf{x};\boldsymbol{\theta})$
minimize the mean square error\citep{Izvekov2005,Noid2008}: 

\begin{align}
\chi^{2}(\boldsymbol{\theta}) & =\left\langle \left\Vert \xi(\mathbf{F}(\mathbf{r}))+\nabla U(\xi(\mathbf{r});\boldsymbol{\theta})\right\Vert ^{2}\right\rangle _{\mathbf{r}}.\label{eq:force-matching-1}
\end{align}

where $\xi(\mathbf{F}(\mathbf{r}))$ denotes the instantaneous all-atom
forces projected onto the CG space, and $\langle\cdot\rangle_{\mathbf{r}}$
is the weighted average over the equilibrium distribution of the atomistic
model, i.e., $\mathbf{r}\sim\mu(\mathbf{r})$.

The effective CG force field $-\nabla U(\mathbf{\mathbf{x}};\boldsymbol{\theta})$
that minimizes $\chi^{2}(\boldsymbol{\theta})$ corresponds to the
mean force\citep{Noid2008}:

\begin{equation}
\mathbf{f}(\mathbf{x})=\langle\xi(\mathbf{F}(\mathbf{r}))\rangle_{\mathbf{r}\mid\mathbf{x}}\label{eq:mean_force}
\end{equation}

where $\mathbf{r}\mid\mathbf{x}$ indicates the equilibrium distribution
of $\mathbf{r}$ constrained to the CG coordinates $\mathbf{x}$,
i.e. the ensemble of all atomistic configurations that map to the
same CG configuration. For this reason, an optimized CG potential
$U(\mathbf{\boldsymbol{x}},\boldsymbol{\theta})$ is also called the
potential of mean force (PMF).

By following statistical estimator theory \citep{Vapnik_IEEE99_StatisticalLearningTheory},
it can also be shown \citep{wang2019} that, the error $\chi^{2}(\boldsymbol{\theta})$
(Eq. \ref{eq:force-matching-1}) can be decomposed into two terms:
\begin{align}
\chi^{2}(\boldsymbol{\theta}) & =\mathrm{PMF}\:\mathrm{error}(\boldsymbol{\theta})+\mathrm{Noise}\label{eq:force-matching-decomp1}
\end{align}
where
\begin{align}
\mathrm{PMF}\:\mathrm{error}(\boldsymbol{\theta}) & =\langle\|\mathbf{f}(\xi(\mathbf{r}))+\nabla U(\xi(\mathbf{r});\boldsymbol{\theta})\|^{2}\rangle_{\mathbf{r}}\nonumber \\
\mathrm{Noise} & =\langle\|\xi(\mathbf{F}(\mathbf{r}))-\mathbf{f}(\xi(\mathbf{r}))\|^{2}\rangle_{\mathbf{r}}.\label{eq:noise}
\end{align}

While the PMF error term depends on the definition of the CG potential
and can be in principle reduced to zero, the noise term does not depend
on the CG potential and it is solely associated with the decrease
in the number of degrees of freedom in the CG mapping, and it is in
general larger than zero. The force matching estimator of Eq. \ref{eq:force-matching-1}
is thus intrinsically very noisy.

\subsection{GDML}

In previous work \citep{wang2019}, we have introduced CGnet to minimize
the error in Eq. \ref{eq:force-matching-1} using a neural network
to parametrize the CG forces. We have demonstrated that the CGnet
approach successfully recovers optimal CG potentials. A large training
dataset enables CGnet to resolve the ambiguity in the coarse-grained
force labels by converging to the respective mean forces. Here, we
explore the Gradient-domain Machine Learning approach (GDML)\citep{ChmielaEtAl_SciAdv17_EnergyConserving,ChmielaEtAl_NatComm18_TowardExact}\textcolor{red}{{}
}as an alternative.

\textcolor{black}{GDML has been used to obtain an accurate reconstruction
of flexible molecular force fields from small reference datasets of
high-level ab initio calculations \citep{ChmielaEtAl_NatComm18_TowardExact,CHMIELA201938,ChmielaEtAl_SciAdv17_EnergyConserving}.
Un}like traditional classical force fields, this approach imposes
no hypothesized interaction pattern for the nuclei and is thus unhindered
in modeling any complex physical phenomena. Instead, GDML imposes
energy conservation as inductive bias, a fundamental property of closed
classical and quantum mechanical systems that does not limit generalization.
This makes highly data efficient reconstruction possible, without
sacrificing generality.

The key idea is to use a Gaussian process ($\mathcal{G}\mathcal{P}$)
to model the force field $\mathbf{f}_{\mathbf{}}$ as a transformation
of an unknown potential energy surface $U$such that

\begin{equation}
\boldsymbol{\mathbf{f}}=-\nabla U\sim\mathcal{G}\mathcal{P}\left[-\nabla\mu_{U}(\mathbf{x}),\nabla_{\mathbf{x}}k_{U}\left(\mathbf{x},\mathbf{x}^{\prime}\right)\nabla_{\mathbf{x}^{\prime}}^{\top}\right].
\end{equation}

Here, $\mu_{U}$ and $k_{U}$ are the mean and covariance functions
of the corresponding energy predictor, respectively.

To help disambiguate physically equivalent inputs, the Cartesian geometries
$\mathbf{x}$ are represented by a descriptor $\mathbf{D}$ with entries:

\begin{equation}
D_{ij}=\left\{ \begin{array}{ll}
{\left\Vert \mathbf{x}_{i}-\mathbf{\mathrm{x}}_{j}\right\Vert ^{-1}} & {\text{ for }i>j}\\
{0} & {\text{ for }i\leq j}
\end{array}\right.
\end{equation}

that introduces roto-translational invariace. Accordingly, the posterior
mean of the GDML model takes the form

\begin{equation}
\hat{\mathbf{f}}(\mathbf{x})=\sum_{i}^{M}\mathbf{J}_{\mathbf{D}(\mathbf{x})}(\nabla_{\mathbf{x}}k_{U}\left(\mathbf{D(x)},\mathbf{D}(\mathbf{x}_{i})\right)\nabla_{\mathbf{x}}^{\top})\mathbf{J}_{\mathbf{D}(\mathbf{x})}^{\top},\label{eq:gdml_force_model}
\end{equation}

where $\mathbf{J}_{\mathbf{D\mathrm{(x)}}}$ is the Jacobian of the
descriptor (see Supplementary Information for details). Due to linearity,
the corresponding expression for the energy predictor can be simply
obtained via (analytic) integration. GDML uses a Matérn kernel $k_{U}(\mathbf{x},\mathbf{x}')$
with restricted differentiability to construct the force field kernel
function

\begin{align}
\boldsymbol{k_{\mathbf{f}}}(\boldsymbol{\mathbf{x}},\boldsymbol{\mathbf{x}}') & =\nabla_{\mathbf{x}}k_{U}\left(\mathbf{x},\mathbf{x}^{\prime}\right)\nabla_{\mathbf{x}^{\prime}}^{\top}\nonumber \\
 & =\left(5\left(\mathbf{x}-\mathbf{x}^{\prime}\right)\left(\mathbf{x}-\mathbf{x}^{\prime}\right)^{\top}-\mathbb{I}\sigma(\sigma+\sqrt{5}d)\right)\label{eq:MaterKernel}\\
 & \cdot\frac{5}{3\sigma^{4}}\exp\left(-\frac{\sqrt{5}d}{\sigma}\right),\nonumber 
\end{align}
\textcolor{black}{where $d=\left\Vert \mathbf{x}-\mathbf{x}^{\prime}\right\Vert $
is the Euclidean distance between the two inputs and $\sigma$ is
an hyperparameter.}

We use this kernel, because empirical evidence indicates that kernels
with limited smoothness yield better predictors, even if the prediction
target is infinitely differentiable. It is generally assumed that
overly smooth priors are detrimental to data efficiency, as the associated
hypothesis space is harder to constrain with a finite number of (potentially
noisy) training examples \citep{rasmussen2004gaussian}. The differentiability
of functions is directly linked to the rate of decay of their spectral
density at high frequencies, which has been shown to play a critical
role in spatial interpolation \citep{stein1999}.

\subsection{Ensemble Learning}

\begin{figure}[h]
\vspace{-0.5cm}
\begin{centering}
\includegraphics[width=1\columnwidth]{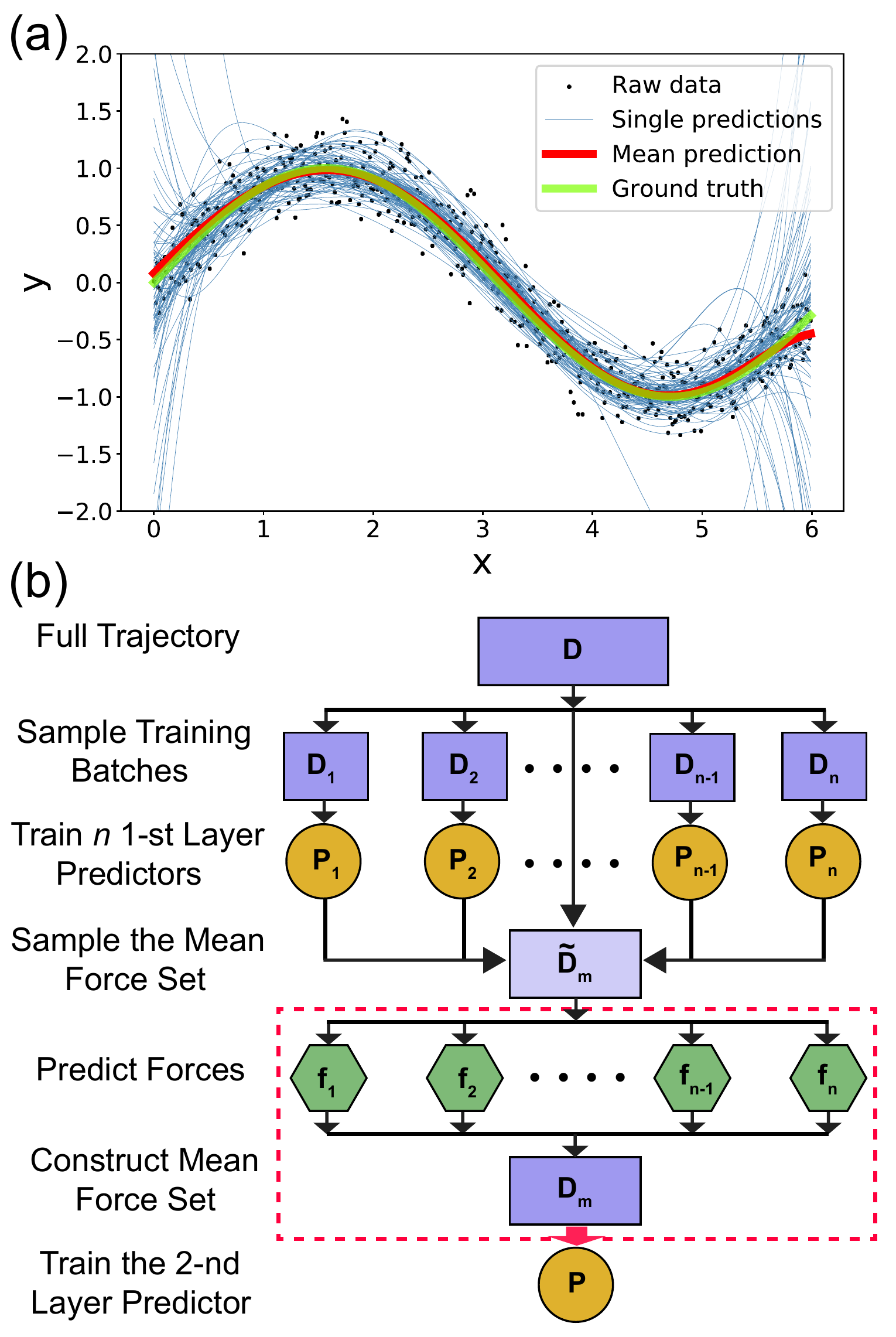}
\par\end{centering}
\caption{\label{fig:EL_schematic-1}Schematic diagram illustrating the principle
of ensemble learning. (a) 1 dimensional toy system. (b). 2-layer training
scheme for learning CG force field using a GDML model. }
\end{figure}

Ensemble learning is a general and widely used machine learning trick
to increase the predictive performance of a trained model by combining
multiple sub-models \citep{Opitz1999,Polikar,Rokach2010,Breiman1996,geman1992neural,Hansen1990,Schapire1990,Kuncheva2003,Bauer1999}.
In this work we use the idea at the basis of a particular ensemble
learning method, called bootstrap-aggregation in the machine learning
literature \citep{Breiman1996}, summarized in the following paragraphs
and Algorithm \ref{alg:2layer-1-1}. This method enables to train
a GDML approach over millions of data points, a task that would be
otherwise impossible. 

In general, we generate a finite set of alternative GDML reconstructions
from randomly drawn subsets of the full MD trajectory and average
them to generate an estimate for the ``expected'' force prediction
at each point. The variability in the individual training sets promotes
flexibility in the structure across all models in the ensemble and
enables us capture the variability in the dataset. We are then able
to compute the expected value for each input by simply taking the
mean of the ensemble. 

Suppose we have a large data set $D:(x,y)$ for training that contains
$N$ samples of pairs of points $x,y$. We would like to train a predictive
model $f$ such that $y=f(x)$, using the data $D$. Instead of training
a single model $f$ using the whole $N$ data points from $D,$ we
first randomly sample $n$ batches: $\{D_{1},D_{2},...,D_{n}\},$
where each batch $D_{i}$ contains $N'$ points. Usually $N$ is too
large to efficiently train a single model, but it is possible to train
sub-models on the different batches $\{f_{1},f_{2},...,f_{n}\}$ if
$N'<<N$. After training all the batches, the final predictive model
$f$ is obtained as the average of all the sub-models:

\begin{equation}
f(x)=\frac{1}{n}\mathop{\sum_{i=1}^{n}f_{i}(x)}.
\end{equation}

This enables us to generate consistent labels for a held-out subset
of the trajectory, which then serves as the basis for another GDML
reconstruction.

We demonstrate how bootstrapping aggregation is used on a simple example,
where we learn an effective curve to fit a one dimensional data set.
As shown in Fig. \ref{fig:EL_schematic-1}a, 600 raw points are uniformly
sampled from $x\in[0,6]$, and the $y$ value of each point is assigned
according to $y_{i}=\sin(x_{i})+0.2\xi$, where $\xi\sim N(0,1)$
is a random noise. These 600 points serve as the noisy training set.
Instead of learning the curve using all 600 points at once, we bootstrap
sample 100 batches from the full data set, where each batch contains
only 20 points. We use a six-order polynomial function to fit 20 points
in each batch, and the 100 fitted curves are shown in blue in Fig.\ref{fig:EL_schematic-1}a.
While each of these 100 blue curves oscillates around the mean and
overfit the data, the mean of the 100 predictors (red curve) is smooth
and agrees with the ground truth $y=\sin(x)$ (green curve) quite
well.We use the idea of ensemble learning to apply GDML to CG problems,
as a 2-layer procedure. Instead of training one single GDML model
using all data, which usually exceed the upper memory limit of GDML,
we train $N$ models $P_{i}|_{i=1}^{N}$ using $N$ data batches,
where each batch contains only $n$ points. In this work, $N=n=1000$.
Since 1000 points is far below the GDML limit, each GDML model $P_{i}$
is easy to train. After obtaining all $N$ GDML models $P_{i}|_{i=1}^{N}$,
we use them to predict the forces $f_{i}$ corresponding to the $i$-th
model for any given CG configuration $\mathbf{x}$ as $f_{i}=P_{i}(\mathbf{x})$.
The mean force (CG force) for a configuration $\mathbf{x}$ is then
the average of the forces for all the models: $f=\frac{1}{N}\sum_{i=1}^{N}f_{i}$
. This average force prediction could be directly used in the CG molecular
simulation but the resulting model would be of low efficiency, since
for each single configuration $\mathbf{x},$ the forces $f_{i}$ need
to be evaluated for all $N$ models $P_{i}|_{i=1}^{N}$ to obtain
an average CG force $f$. 

This low evaluation efficiency motivates us to propose a 2-layer procedure
to speed up the evaluation of the mean force prediction. We generate
a new batch of data $\tilde{D}_{m}$ which contains $n'$ points ($n'=3000$
in this work). For all CG configurations in $\tilde{D}_{m}$, we use
$N$ predictors to evaluate their forces and compute the mean forces.
This produces a new data set $D_{m}$ where the $n'$ configurations
are associated to the corresponding mean forces. Constructing $D_{m}$
can be fast because the mean forces are computed only for $n'$ points,
usually a few thousands. 

\begin{algorithm}
\textbf{2Layer}-GDML($D,N,n,n'$) 
\begin{enumerate}
\item Sample $N$ data batches from the original bulk data set $D:\{D_{1},D_{2},...,D_{N}\}$,
each data batch $D_{i}$ contains $n$ randomly sampled points, each
data point $d=(r,f)$ includes a molecular configuration part $r$
and a force part $f$
\item Sample one additional data batch $\tilde{D}_{m}$ from the original
bulk data set $D$, that contains $n'$ points, $d_{i}=(r_{i},f_{i})|_{i=1,2,...,n'}$,
each points $d_{i}$ also includes a molecular configuration part
$r_{i}$ and a force part $f_{i}$, and $j$ indicates the point index
in data batch $\tilde{D}_{m}$
\item For $i=1,...,N:$ \# Loop over $N$ batches:
\begin{enumerate}
\item Train GDML model $P_{i}$ using data batch $D_{i}$
\item Predict forces for all $n'$ configurations $r_{j}|_{j=1,2,...n'}$
in data batch $\tilde{D}_{m}$ using model $P_{i}$, which is denoted
as $f_{j}^{i}|_{j=1,2,...,n'}$
\end{enumerate}
\item Construct the mean force set $D_{m}$, which also includes $n'$ points,
the configuration part $r_{j}$ for each point is the same as $\tilde{D}_{m}$,
but the force part $f_{j}$ is the averaged force computed using the
$N$ GDML models: $f_{j}=\frac{1}{N}\sum_{i=1}^{N}f_{j}^{i}$
\item Train the 2nd-layer model $P$ using the constructed mean force set
$D_{m}$
\end{enumerate}
\caption{\label{alg:2layer-1-1}2-layer training scheme}
\end{algorithm}

Once the mean force set $D_{m}$ is obtained, we can train a single
final model $P$ using the entire $D_{m}$. Since $D_{m}$ contains
the mean forces, the final model $P$ also predicts the mean forces
for the CG configurations. $P$ is easy to train due to small size
of $D_{m}$ ($n'$ is far below the GDML limit) and the force evaluation
for the final model $P$ is much more efficient than by evaluating
$N$ models $P_{i}|_{i=1}^{N}$. The general procedure of the 2-layer
scheme is illustrated in Algorithm \ref{alg:2layer-1-1}.

The hyperparameters that control the performance of the final model
are the two kernel sizes $\sigma_{1},\sigma_{2}$, for each layer
(see Eq. \ref{eq:MaterKernel}).\textcolor{blue}{{} }Another hyperparameter
is the regularization coefficient of the ridge term, and is set to
the standard value ($\lambda=1\times10^{-15}$) as in the original
GDML paper \citep{ChmielaEtAl_SciAdv17_EnergyConserving}. We conduct
a 2D cross-validation search to determine $\sigma_{1}$ and $\sigma_{2}$.
The algorithm for the cross-validation of the ensemble learning GDML
is shown in Algorithm \ref{alg:2layer_CV-1-1}. The parameters $N,n,n',K$
are selected as $N=n=1000,n'=3000,K=5,$ $K$ is the number of folds
for the cross-validation, and the total number of points in $D$ is
$1,000,000$.

\begin{algorithm}
\textbf{2Layer}-GDML-CV($D,N,n,n',K$) 
\begin{enumerate}
\item Sample $N$ data batches from the original bulk data set $D:\{D_{1},D_{2},...,D_{N}\}$,
each data batch $D_{i}$ contains $n$ randomly sampled points, each
points $d=(r,f)$ includes the molecular configuration part $r$ and
the force part $f$
\item Sample one additional data batch $\tilde{D}_{m}$ from the original
bulk data set $D$, that contains $n'$ points, $d_{i}=(r_{i},f_{i})|_{i=1,2,...,n'}$,
each points $d_{i}$ also includes the molecular configuration part
$r_{i}$ and the force part $f_{i}$, and $j$ indicates the point
index in data batch $\tilde{D}_{m}$
\item For $i=1,...,N:$ \# Loop over $N$ batches:
\begin{enumerate}
\item Train GDML model $P_{i}$ using data batch $D_{i}$
\item Predict forces for all $n'$ configurations $r_{j}|_{j=1,2,...n'}$
in data batch $\tilde{D}_{m}$ using model $P_{i}$, which is denoted
as $f_{j}^{i}|_{j=1,2,...,n'}$
\end{enumerate}
\item Divide $N$ data batches into $K$ subsets of batches: $SD=\{SD_{1},SD_{2},...,SD_{K}\}$,
where each subset $SD_{i}$ contains $N/K$ batches.
\item For the $j^{th}$ point in $\tilde{D}_{m}$, divide $N$ of its predicted
forces $f_{j}^{i}|_{i=1,...,N}$ into subsets $SF_{j}=\{SF_{j,1},SF_{j,2},...,SF_{j,K}\}$,
where each subset contains $N/K$ force tags that are consistent with
the division in step 4, and $j=1,2,...,n'$
\item For $l=1,...,K:$ Loop over $K$ cross-validation folds:
\begin{enumerate}
\item For the $j^{th}$ point in $\tilde{D}_{m}$, compute the mean forces
using all forces from $SF_{j}\backslash SF_{j,l}$ (excluding $SF_{j,l})$,
where $j=1,2,...,n'$, after obtaining the mean forces for all $n'$
configurations in $\tilde{D}_{m}$, construct the $l^{th}$ mean force
set $D_{m,l}$
\item Train the $l^{th}$ second layer model $P2_{l}$ using $D_{m,l}$
\item Compute the validation error of model $P2_{l}$ using all data points
from the excluded set $SD_{l}$, and denote the error as $E_{l}$
\end{enumerate}
\item Return cross-validation score $\frac{1}{K}\sum_{k=1}^{K}E_{k}$
\end{enumerate}
\caption{\label{alg:2layer_CV-1-1}2-layer training scheme with cross-validation}
\end{algorithm}

\subsection{Stratified Sampling }

\begin{figure}[h]
\begin{centering}
\includegraphics[width=1\columnwidth]{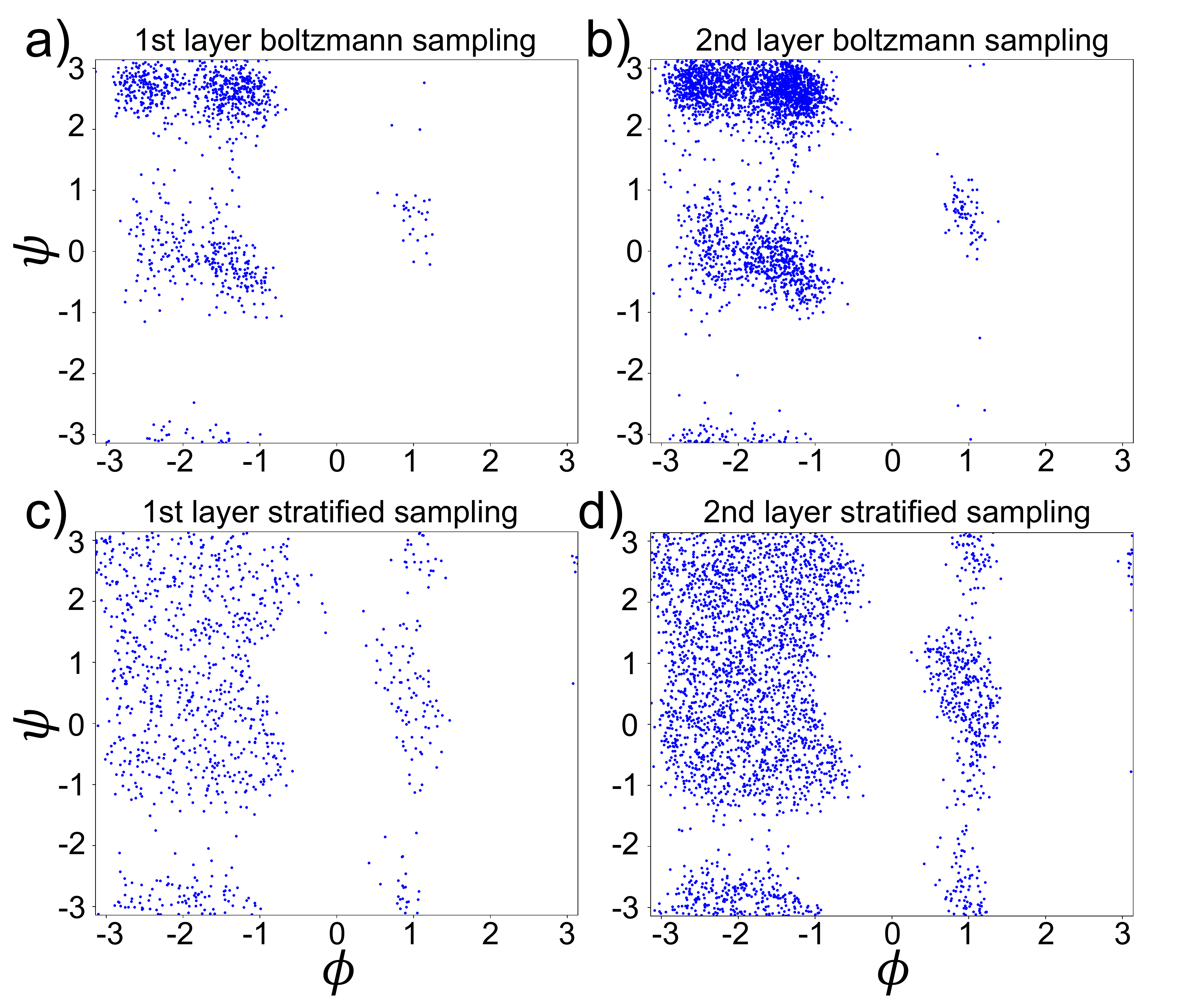}
\par\end{centering}
\caption{\label{fig:Stratigic_sampling}Stratified sampling of the training
set for alanine-dipeptide in the dihedral angles $(\psi,\phi)$ space.
(a) Regular (Boltzmann distributed) sampling of 1000 points for the
1st-layer. (b) Regular sampling of 3000 points for the 2nd-layer.
(c) Uniformly stratified sampling of 1000 points in the $(\psi,\phi)$
space for the 1$^{st}$ layer. (d) Uniformly stratified sampling of
3000 points in the $(\psi,\phi)$ space for the 2$^{nd}$ layer. }
\end{figure}

Another crucial factor that impacts the overall performance of our
machine learning model is the distribution of the training data. As
our training data are obtained from extensive MD simulations, they
are distributed according to the Boltzmann distribution in the molecule
configuration space. If a small batch of data is randomly sampled
from the whole data set, the large majority of the data will reside
in low free energy regions, while data in high free energy region,
such as transition barriers, are underrepresented. Fig. \ref{fig:Stratigic_sampling}a
shows that, in the case of alanine dipeptide, most of the data in
a small batch of randomly selected points are located in the free
energy minima on the left side of the $(\phi,\psi)$ dihedral angle
space. If batches from this biased distribution are used in the ensemble
learning, the errors for predicting the PMF in high free energy regions
would be very large, because the models will not be trained efficiently
in these sparse data regions.

In order to solve this issue, we sample the data for the batches uniformly
in the $(\phi,\psi)$ dihedral angles space of alanine dipeptide,
as shown in Fig. \ref{fig:Stratigic_sampling}b. In this way, all
relevant regions in the configurational space are equally represented
in the training set, including transition states. The advantage of
this strategic sampling is illustrated in more details in section
\ref{sec:Results}. 

\begin{figure*}[!tp]
\begin{centering}
\includegraphics[width=1\textwidth]{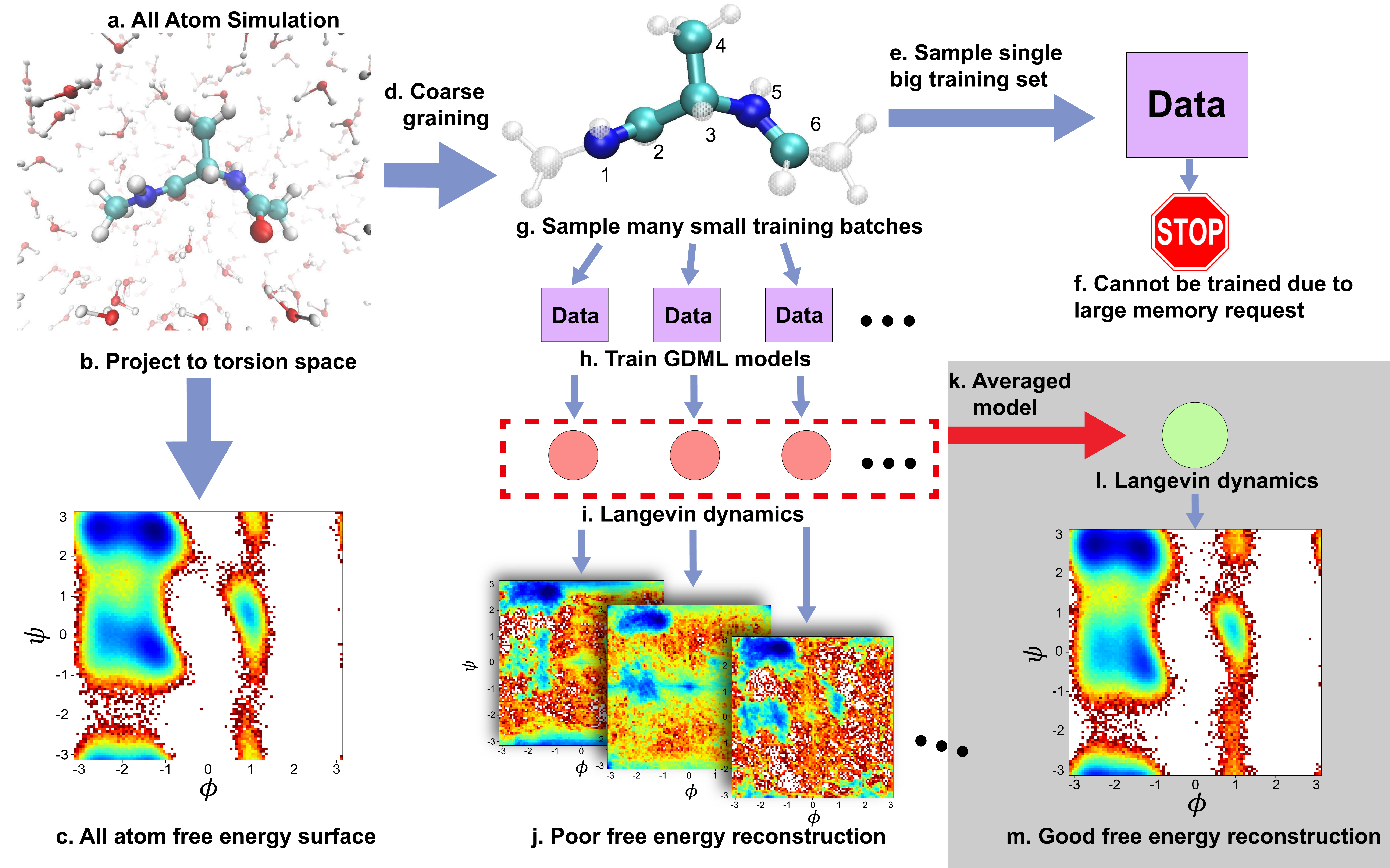}
\par\end{centering}
\centering{}\caption{\label{fig:pipeline} Pipeline of learning the CG forcefield with
the GDML model. (a) All atom simulation of alanine dipeptide in water.
(b) We compute the two dihedral angles $\phi$ and $\psi$, project
the simulation data on to the $(\phi,\psi)$ space. (c) All-atom free
energy surface in $(\phi,\psi)$ space. (d) The coarse graining model
contains only six heavy atoms from the original molecule. (e) We could
sample enough point for training a CG model, the data set is usually
big. (f) Training the GDML model with one big data set requires large
memory, which hinder the application of GDML to coarse graining a
molecule. (g) Instead of sampling one big training set, we sample
many smaller training sets. (h) We train GDML models with each small
training set. (i) We use Langevin dynamics to simulate a CG MD trajectory
with each trained GDML model. (j) Similar to (b) and (c), we compute
the free enenrgy surface in $(\phi,\psi)$ space for each trajectory,
and we find that these single models poorly recovered the correct
free energy surface. (k) We can obtain an extra model which is the
average of all models we trained in step (h). The averaging proceedure
indicated by the red dashed box and arrow corresponds to the red box
and arrow in Figure. \ref{fig:EL_schematic-1}(b). (l) We can simulate
the averaged model using Lagevin dynamics. (m) The average CG model
can correctly reconstruct the free energy surface of the molecule.
The final result is highlighted in a light gray box.}
\end{figure*}

\subsection{Simulating the CG-GDML Model}

After training the 2-layer GDML model, we use an over-damped Langevin
dynamics to generate a trajectory and sample the CG potential $U(\mathbf{x};\theta)$:

\begin{equation}
\mathbf{x}_{t+\tau}=\mathbf{x}_{t}-\tau\frac{D}{k_{B}T}\nabla U(\mathbf{x}_{t})+\sqrt{2\tau D}\boldsymbol{\xi}\label{eq:Smoluchowski}
\end{equation}

where $\mathbf{x}_{t}$ ($\mathbf{x}_{t+\tau}$) is the CG configuration
at time $t$ ($t+\tau$), $\tau$ is the time step, $D$ is the diffusion
constant, and $\xi$ is a vector of independent Gaussian random variables
with zero-mean and identity covariance matrix (Wiener process). To
sample the trained potential more efficiently, we generate 100 independent
trajectories in parallel, with initial configurations randomly sampled
from the original data set.

\subsection{Including Physical Constraints}

When an over-damped Langevin dynamics (Eq. \ref{eq:Smoluchowski})
is used to generate a trajectory with a CG potential trained on a
finite dataset, one undesired situation may happen: since the dynamics
is stochastic, there is a chance that the simulated CG trajectory
may diffuse away from the domain of the data used in the training,
generating unphysical configurations. For example, the stretching
of a bond too far away from the equilibrium distance is associated
with a very high energy cost and is never observed in simulation with
a force field at finite temperature. In simulation with a machine-learned
CG potential, there is no mechanism for preventing such as unphysical
bond-stretching. Similarly to what we proposed in our recent work
\citep{wang2019}, this problem can be solved by including a prior
potential energy $U_{prior}(\mathbf{x})$ incorporating physical prior
knowledge on the system:

\[
U(\mathbf{x};\boldsymbol{\theta})=U_{diff}(\mathbf{x;\boldsymbol{\theta}})+U_{\mathrm{prior}}(\mathbf{x})
\]

where $U_{prior}(\mathbf{x})$ has harmonic terms modeling bond and
angle stretching, with parameters extracted from the training data
by Boltzmann inversion. $U_{diff}(\mathbf{x};\boldsymbol{\theta})$
is the difference between the total CG potential and $U_{prior}(\mathbf{x})$.
The forces obey a similar relation:

\[
\begin{array}{cc}
-\nabla U(\mathbf{x};\boldsymbol{\theta})=-\nabla U_{diff}(\mathbf{x;\boldsymbol{\theta}})-\nabla U_{\mathrm{prior}}(\mathbf{x})\end{array}
\]

so the loss function of the model becomes:

\begin{multline}
\chi^{2}(\boldsymbol{\theta})=\Bigl\langle\Bigl\Vert\xi(\mathbf{F}(\mathbf{r}))-\nabla U_{prior}(\mathbf{x})+\\
+\nabla U_{diff}(\xi(\mathbf{r});\boldsymbol{\theta})\Bigr\Vert^{2}\Bigr\rangle_{r}.\label{eq:}
\end{multline}

Differently from what was done in a neural network model \citep{wang2019},
the prior potential is not added directly to the trained model: the
prior forces are first evaluated and subtracted from the all-atom
forces, and the GDML is trained over this force difference. Once the
model is trained, the total energy (and forces) is obtained by adding
back the prior energy (and forces) to the one obtained from the trained
model. 

\section{Results\label{sec:Results}}

We illustrate the results of the approach discussed above on a simple
molecular system, namely the coarse-graining of the alanine-dipeptide
molecule from the atomistic model in explicit water into a 6-bead
CG model. The all-atom model of alanine dipeptide consists of 22 atoms
and 651 water molecules, for a total of a few thousand degrees of
freedom. As illustrated in Fig. \ref{fig:pipeline}, for the CG representation
we select the 5 central backbone atoms of the molecule, with additionally
a 6$^{th}$ atom to break the symmetry and differentiate right- or
left- handed representations. The overall pipeline for the coarse
graining and the training procedure that is discussed below is also
summarized in Figure \ref{fig:pipeline}.

\begin{figure*}[!tp]
\begin{centering}
\includegraphics[width=1\textwidth]{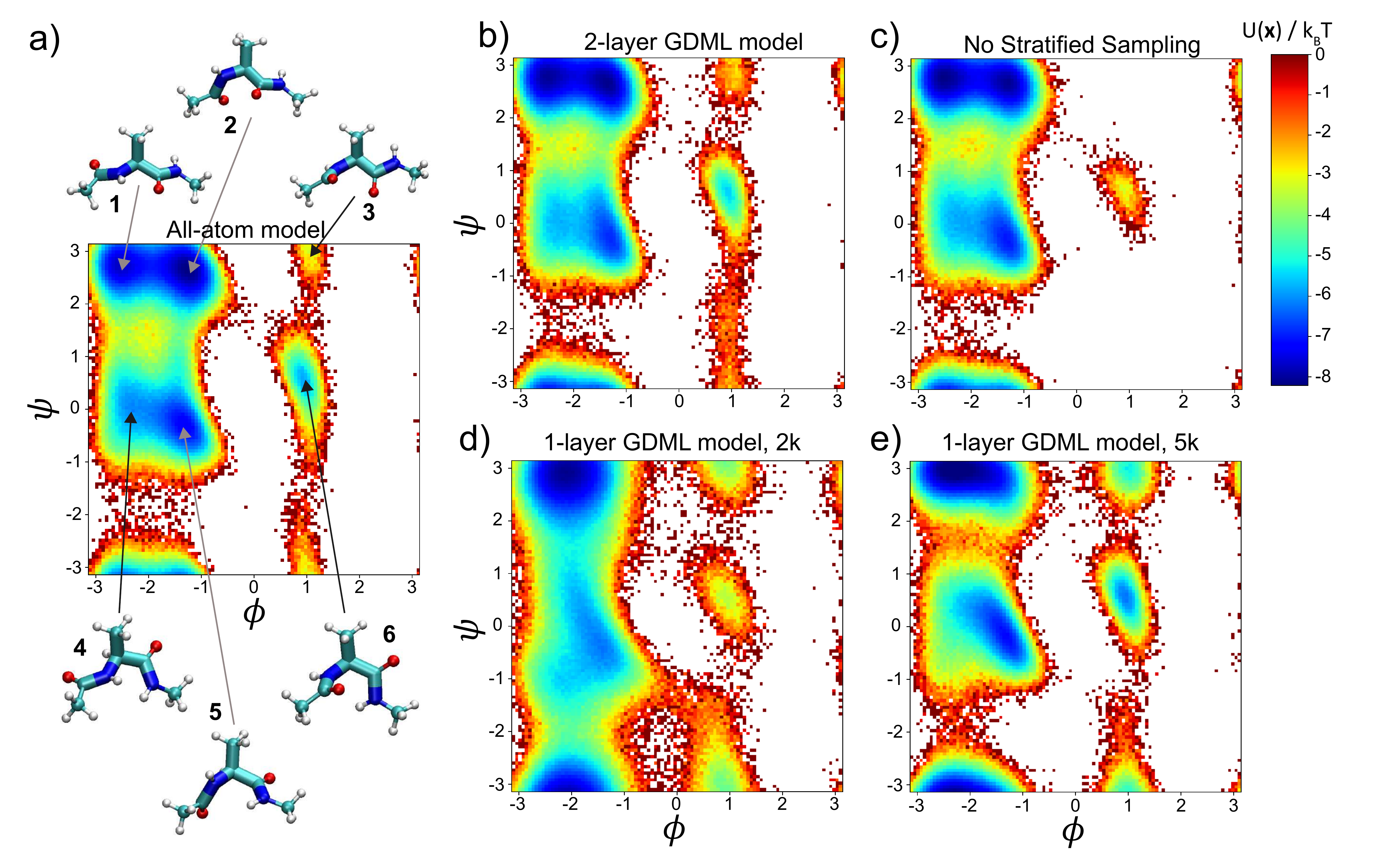}
\par\end{centering}
\centering{}\caption{\label{fig:dialanine_FES}Free energy surface in $(\psi,\phi)$ space
for the trained GDML models. (a) reference (all-atom) free energy
landscape, and all representative configurations of the molecule in
the six minima sampled from the all-atom trajectory (CPK representation)
and from a CG simulation with the 2-layer GDML model (thick bonds).
(b) Free energy landscape from the 2-layer GDML model. (c) Free energy
from a 2-layer GDML model with no stratified sampling. (d) Free energy
from a traditional single-layer GDML model, trained with 2000 points.
(e) Free energy from a traditional single-layer GDML model, trained
with 5000 points. }
\end{figure*}

We compute the free energy of the alanine dipeptide as a function
of the two dihedral angles $\phi,\psi$, where $\phi$ is defined
by atoms 1, 2, 3, 5, and $\psi$ by atoms 2, 3, 5, 6 (see Fig. \ref{fig:pipeline}).
As shown in Fig. \ref{fig:dialanine_FES}a, there are six metastable
states in the free energy landscape of the all-atom model of alanine
dipeptide. Fig.\ref{fig:dialanine_FES}b shows that the final 2-layer
GDML CG model correctly reproduces the free energy landscape of alanine
dipeptide: the free energy obtained from the trajectories (generated
by numerical integration of Eq.\ref{eq:Smoluchowski}) of the CG model
also exhibits six minima, with depths close to the ones of the corresponding
minima in the all-atom model. Representative configurations from the
six metastable states are shown in Fig.\ref{fig:dialanine_FES}a for
the all-atom (CPK representation) and the CG model (thick bond representation).
Moreover, as shown in SI Fig. \ref{fig:angle_bond_distribution-1},
the bond and angle distribution from the CG simulation are also consistent
with the all atom simulation.

The GDML model shown in \ref{fig:dialanine_FES}b is optimized based
on the minimum cross-validation error over a 2-dimensional grid, spanned
by the parameters $\sigma_{1}$ and $\sigma_{2}$, which are the kernel
width for the 1$^{st}$ and the 2$^{nd}$ layer models. We find that
the values $(\sigma_{1},\sigma_{2})=(100,10)$ give the smallest cross-validation
error. Details on the cross-validation search can be found in SI Fig.
\ref{fig:Cross_validation-1}. 

Fig. \ref{fig:dialanine_FES}c reports the free energy landscape corresponding
to a CG model obtained with a 2-layer GDML but where the selection
of the data for the sub-model is performed according to the Boltzmann
distribution (that is, uniform sampling along the MD trajectory) instead
of the stratified sampling scheme discussed above (uniform sampling
in the $\phi,\psi$ space). While the free energy around the region
of the deepest free energy minima in the $\phi,\psi$ space is quite
accurate, the lowly populated metastable state (indicated as state
3 in \ref{fig:dialanine_FES}a) is completely missing in \ref{fig:dialanine_FES}c,
because of the scarcity of training points in this region. 

As a comparison, Fig. \ref{fig:dialanine_FES}d shows the results
when a single-layer GDML model is trained on only 2000 points. Although
this model identifies the general location of the metastable states,
the free energy landscape is significantly distorted with respect
to the all-atom one. This poor reconstruction performance is due to
the limited size of the training set, which is not extensive enough
to enable a stable estimate of the expected forces for the reduced
representation of the input. We also trained a single-layer GDML model
on 5000 points. As shown in Fig. \ref{fig:dialanine_FES} e, the free
energy of this model presents a slightly improvement with respect
Fig. \ref{fig:dialanine_FES}d because of the increased number of
training points. However, the overall quality is still low comparing
to the atomistic model. We expect the reconstructed free energy to
improve further if we trained a model using much more data, but this
is hindered by the memory requirement: it requires about 160 GB memory
to train a model with 5000 points, which is almost at the upper limit
of our computational ability.

To quantify the performance of the different approaches, we compute
the mean square error (MSE) of the free energy difference of the different
CG models compared to the atomistic model (Fig. \ref{fig:dialanine_FES}a,
Table \ref{tab:Free-energy-difference}, see \citep{wang2019} for
details). As expected, the 2-layer GDML model has the smallest free
energy MSE, which is about $0.363\pm0.112\,(k_{B}T)^{2}$, when it
is trained with all 1000 batches. The single layer GDML gives the
largest free energy difference ( $2.947\,(k_{B}T)^{2}$ if trained
with 2000 points, and $1.641\,(k_{B}T)^{2}$ if trained with 5000
points). If no stratified sampling is used, the free energy difference
is $0.861\,(k_{B}T)^{2}$, and most of this value is due to the discrepancies
in the free energy $\phi>0$ region. 

As a baseline, we also compute the free energy difference obtained
by a CG model designed by means of a neural network, CGnet \citep{wang2019}.
Previously, we have applied CGnet to alanine-dipeptide, but it was
a model based on a 5 atom CG scheme, and we included two dihedral
angles as input features to break the symmetry. To make the CGnet
model consistent with the CG scheme used in this work, we modified
it to contain 6 atoms (as in the GDML model), and no dihedral angles
features were included (only distances are used as input). This CGnet
model is trained with the same number of points as the GDML model
(i.e. 1,000,000 points from 1000 batches). The resulting CGnet free
energy MSE is $0.475\pm0.103\,(k_{B}T)^{2}$, a value slightly larger
than the 2-layer GDML model. This result shows that the accuracy of
a kernel approach can indeed be comparable to or even better than
a neural network approach on the same system. 

\begin{figure}[h]
\begin{centering}
\includegraphics[width=0.85\columnwidth]{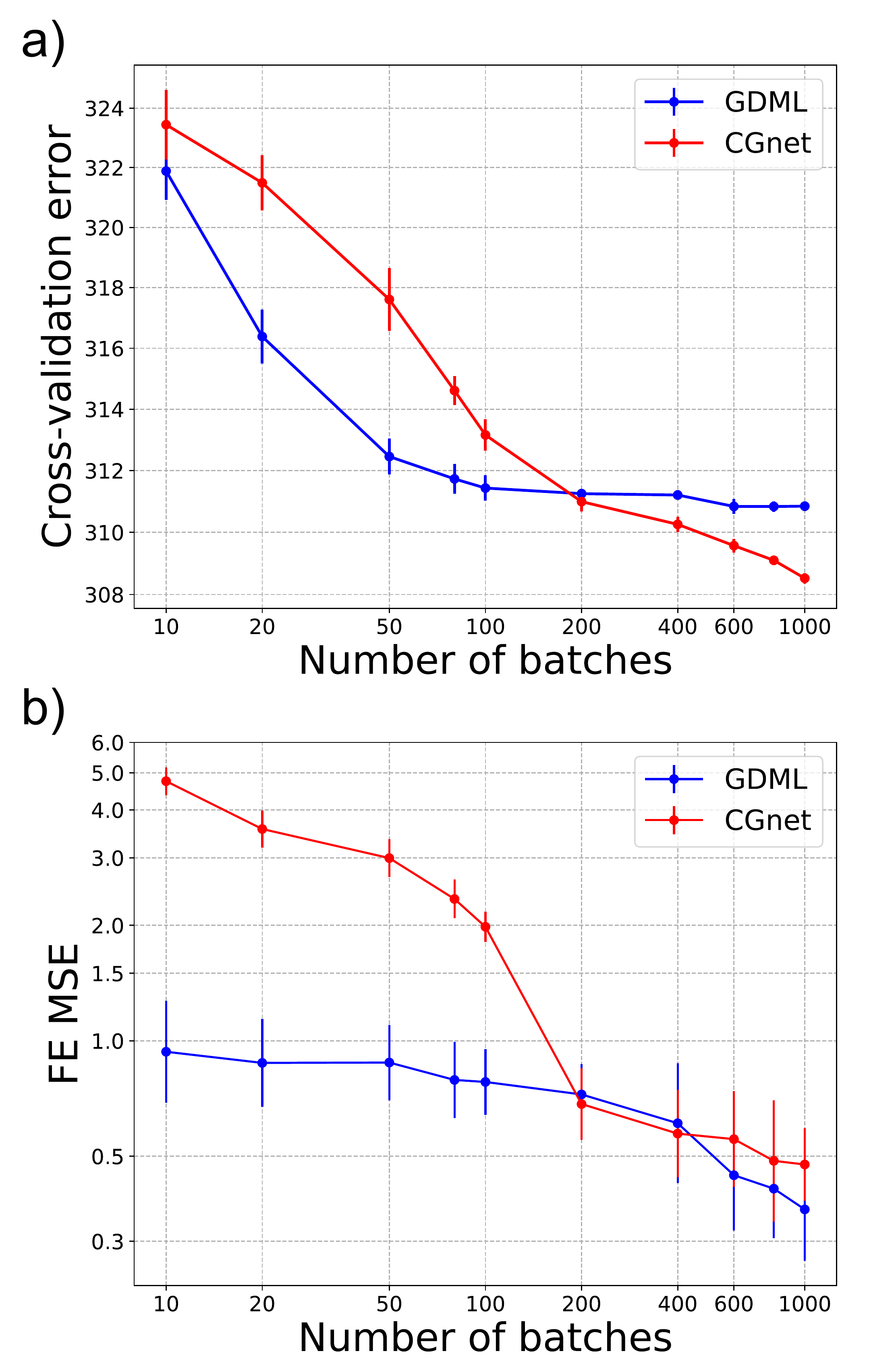}
\par\end{centering}
\caption{\label{fig:CVbatchNumber-1}Cross-validation error (a) and free energy
mean square error (MSE) (b) as a function of the number of batches. For
CGnet, the training set size is equal to $1000\times$ number of batches.
The units for the crosss-validation error are $kcal/mol^2/\text{\AA}^2$, while the
units for the free energy MSE are $(k_BT)^2$.}
\end{figure}

\begin{figure}[h]
\begin{centering}
\includegraphics[width=1\columnwidth]{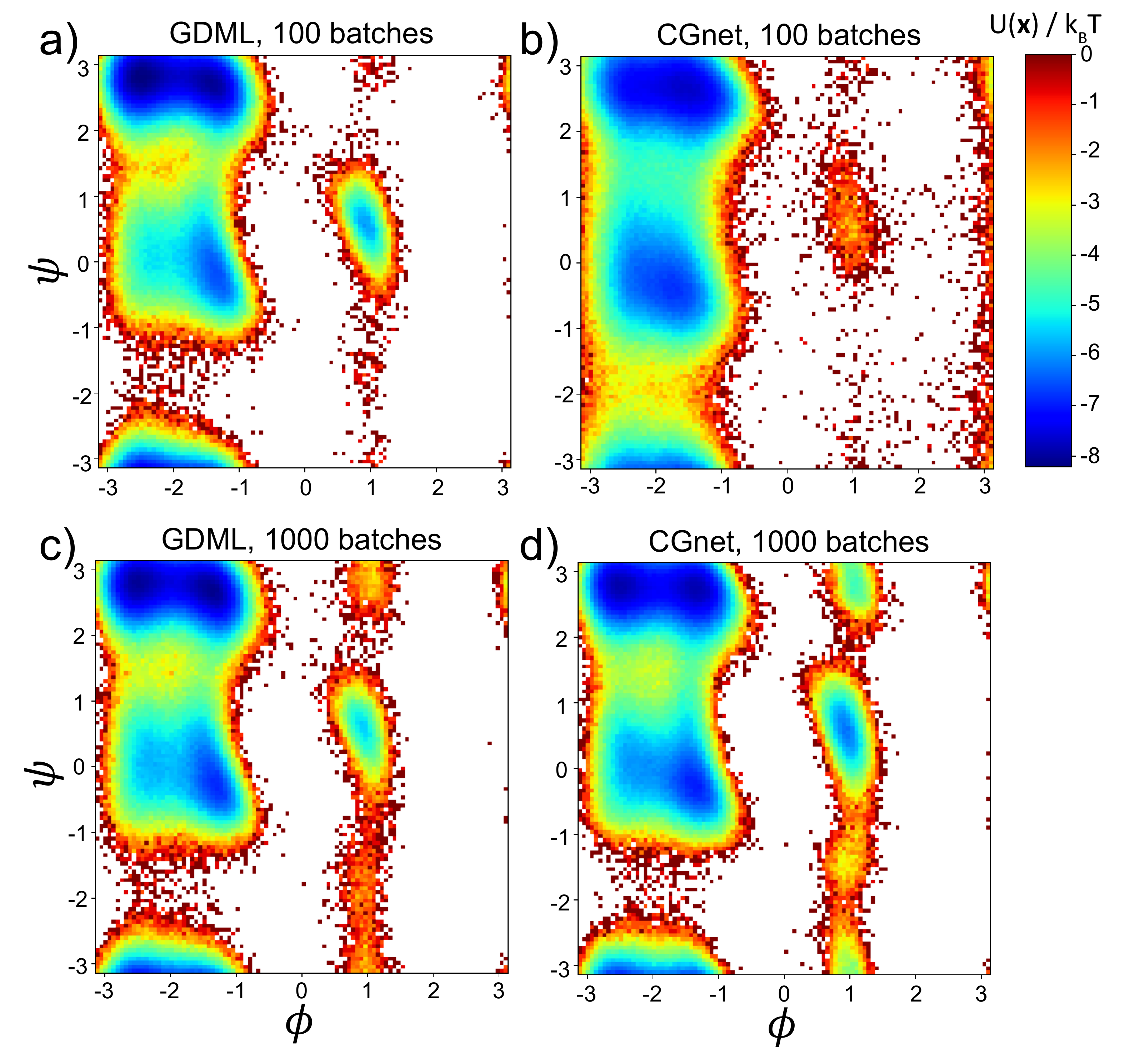}
\par\end{centering}
\caption{\label{fig:FES_size-1}Free energy as a function of the alanine dipeptide
dihedral angles, for a 2-layer GDML CG model with number of batches
$N_{Batch}=100$ (a) and $N_{Batch}=1000$ (c), and for CGnet with
$N_{Batch}=100$ (b) and $N_{Batch}=1000$ (d).}
\end{figure}

We have also investigated the effect of the batch number (or the training
set size). We computed the cross-validation error with different training
set size, from $10$ to $1000$ batches for the GDML model, or equivalentely
from $10,000$ to $1,000,000$ points for CGnet. Fig. \ref{fig:CVbatchNumber-1}a
shows that as the batch number increases from $10$ to $1000$, the
cross-validation error for GDML model drops quickly, and reaches convergence
with a batch number $>600$ . The cross-validation error for CGnet
is significantly larger than for the GDML model when the number of
batches (or, equivalently, the training set size) is small. When the
batch number is larger than 200, the cross-validation error for CGnet
becomes smaller than for GDML. Similarly, if we compare the free energy
MSEs, as shown in Fig. \ref{fig:CVbatchNumber-1}b, the free energy
constructed by GDML with a small training set is significantly better
than the corresponding free energy constructed by CGnet. On the other
hand, with a large training set, the MSEs are comparable to each other.
Typical free energy profiles are shown in Fig. \ref{fig:FES_size-1}a-d,
and their corresponding MSE values are shown Table \ref{tab:Free-energy-difference}.
These results show that with enough data, the 2-layer GDML model and
CGnet perform similarly well. However, the 2-layer GDML model is more
data efficient and has a better ability to extrapolate the force prediction
to unsampled configurations, thus outperforming CGnet for small training
sets.

\begin{table*}[!tp]
\caption{\label{tab:Free-energy-difference}Free energy mean square error (MSE)
comparison for different CG models trained with different number of
points, which is in the unit of thousand (k). The error is computed
as the mean square error of the free energy of alanine-dipeptide in
($\psi,\phi$) space, relative to the atomistic free energy \citep{wang2019}.
The MSE values are in units of $(k_{B}T)^{2}$.}

\centering{}%
\begin{tabular}{|c|c|c|c|c|}
\hline 
\multirow{2}{*}{Model} & \multirow{2}{*}{100 k} & \multirow{2}{*}{1000 k} & \multirow{2}{*}{2 k} & \multirow{2}{*}{5 k}\tabularnewline
 &  &  &  & \tabularnewline
\hline 
\hline 
CGnet & $1.982\pm0.181$ & $0.475\pm0.103$ & - & -\tabularnewline
\hline 
2-layer GDML & \textbf{$0.781\pm0.154$} & \textbf{$0.363\pm0.112$} & - & -\tabularnewline
\hline 
Boltz. Samp. 2-layer GDML & - & $0.861\pm0.167$ & - & -\tabularnewline
\hline 
1-layer GDML  & - & - & $2.947\pm0.264$ & \textbf{$1.641\pm0.243$}\tabularnewline
\hline 
\end{tabular}
\end{table*}

\section{Conclusions}

In this work we combine the idea of ensemble learning with GDML, to
apply it to the coarse graining problem. GDML is a kernel method to
learn molecular force fields from data, and allows to model nonlinearity
and multi-body effects without the need of providing a functional
form for the potential. The GMDL approach was originally proposed
to learn molecular forces from quantum simulation data. When quantum
calculations are used, the error on the force matching loss could
in in principle be zero, and a few thousand points are enough to construct
and build an accurate, smooth and conserved force field. However,
when applied to coarse-graining , the force matching loss contains
a non zero term due to the dimensionality reduction and the learning
problem becomes very noisy. For this reason a lot more data points
are needed from atomistic simulations to learn a CG potential of mean
force. The large amount of input data would presently hinder the application
of GDML to the CG problem. In order to circumvent this problem, we
use ensemble learning. The basic idea consists in breaking down the
learning problem into small batches, that can be more easily solved,
and combine the resulting different models into a final solution.
Following this approach, we do not train one single GDML model using
all the data, but propose a 2-layer training scheme: in the first
layer, we generate $N$ data batches, each containing a number of
points far below the GDML limit. $N$ models are trained on the different
batches and are combined into a final model by taking the average.
We show that the prediction of the CG 2-layer model accurately reproduces
the thermodynamics of the atomistic model. 

Consistently with previous work \citep{wang2019}, we show that, when
applying machine learning methods to design force fields for molecular
systems, the addition of physical constraints enforce proper asymptotic
of the model. In the design of CG potentials, physical constraints
can be introduced by means of a prior potential energy term that prevent
the appearance of spurious effects in non-physical regions of the
configurational landscape.

A good GDML model should be able to construct a smooth and globally
connected conserved force field. However, when the 2-layer approach
is used some of the molecular configurations with high free energy
are poorly sampled in the training set, introducing large errors in
the resulting model. In order to solve this problem, we sample the
data uniformly in the low dimensional space defined by two collective
coordinates rather than uniformly from the simulation time series.
In the example of alanine dipeptide discussed here, the dihedral angles
$\phi,\psi$ are chosen as collective coordinates. 

In our previous work, we proposed CGnet \citep{wang2019}, a neural
network approach to design CG models. The overall free energy reconstruction
obtained with the GDML model is comparably accurate as what was obtained
with CGnet when the training set size is sufficiently large. However,
the GDML model is significantely more accurate when the training set
size is small, indicating that a kernel approach is data-efficient
and could in principle provide more accurate CG models especially
with small training sets.

However, there are still several challenges in order to apply GDML
for the coarse-graining of macromolecular systems. In larger systems,
a more general definition is needed for the collective coordinates
defining the low dimensional space for the uniform sampling of the
training batches. These collective coordinates could in principle
be extracted from the trajectory data \citep{RohrdanzEtAl_AnnRevPhysChem13_MountainPasses,NoeClementi_COSB17_SlowCVs},
for instance by means of time-lagged Independent Component Analysis
(tICA)\citep{Perez-Hernandez2013,SchwantesPande_JCTC13_TICA,ZieheMueller_ICANN98_TDSEP,Belouchrani1997,Molgedey_94},
kernel PCA\citep{Klaus1997,Klaus1998,Klaus2001} or diffusion maps
\citep{RohrdanzClementi_JCP134_DiffMaps}. 

The decomposition of the large input data set into an ensemble of
small batches has been used here to solve memory issues when training
a GDML model. However, the computation is still expensive and we expect
it to become even more expensive as the size of the molecular system
increases. As the number of data batches and batch size grow, the
$\text{{Nyström}}$ approximation of the kernel or other numerical
approaches may be a promising solution to increase the computational
efficiency. 

As for the neural network model, the GDML model trained by force matching
can capture the thermodynamics of the system, but there is no guarantee
that the dynamics is also preserved. Alternative approaches need to
be defined to solve this problem \citep{Nuske2019}. 

Finally, the GDML model presented here is trained on a specific molecule,
and it is not directly transferable to different systems. Ultimately,
a transferable CG model would be needed for the general application
to large systems that can not be simulated by atomistic simulations.
The trade-off between accuracy and transferability in CG models is
an open research question that we will investigate in future work.

\section{Supplementary Material}
See Supplementary Material for more details about the hyperparameter search, a
discussion on the prior energy, and more information on the descriptors used in
the GDML.

\begin{acknowledgments}
We thank Eugen Hruska and Feliks Nüske for comments on the manuscript.
This work was supported by the National Science Foundation (CHE-1738990,
CHE-1900374, and PHY-1427654), the Welch Foundation (C-1570), the
MATH+ excellence cluster (AA1-6, EF1-2), the Deutsche Forschungsgemeinschaft
(SFB 1114/A04), the European Commission (ERC CoG 772230 ``ScaleCell'')
and the Einstein Foundation Berlin (Einstein Visiting Fellowship to
CC). Simulations have been performed on the computer clusters of the
Center for Research Computing at Rice University, supported in part
by the Big-Data Private-Cloud Research Cyberinfrastructure MRI-award
(NSF grant CNS-1338099), and on the clusters of the Department of
Mathematics and Computer Science at Freie Universität, Berlin. K.-R.M.
acknowledges partial financial support by the German Ministry for
Education and Research (BMBF) under Grants 01IS14013A-E, 01IS18025A,
01IS18037A, 01GQ1115 and 01GQ0850; Deutsche Forschungsgesellschaft
(DFG) under Grant Math+, EXC 2046/1, Project ID 390685689 and by the
Technology Promotion (IITP) grant funded by the Korea government (No.
2017-0-00451, No. 2017-0-01779). S.C. acknowledges the BASLEARN -
TU Berlin/BASF Joint Laboratory, co-financed by TU Berlin and BASF
SE.
Part of this research was performed while the authors were visiting the Institute
for Pure and Applied Mathematics (IPAM), which is supported by the National
Science Foundation (Grant No. DMS-1440415).
\end{acknowledgments}

The data that support the findings of this study are available from the
corresponding author upon reasonable request.

\bibliographystyle{unsrt}

\end{document}